# A STUDY ON THE RELATIONSHIP BETWEEN DEPTH MAP QUALITY AND THE OVERALL 3D VIDEO QUALITY OF EXPERIENCE

*Amin Banitalebi-Dehkordi[1], Student Member, IEEE, Mahsa T. Pourazad[1,2], Member, IEEE, and Panos Nasiopoulos[1], Senior Member, IEEE*

[1]Department of Electrical and Computer Engineering, University of British Columbia, Canada
[2]TELUS Communications Inc. Canada

**ABSTRACT**

The emergence of multiview displays has made the need for synthesizing virtual views more pronounced, since it is not practical to capture all of the possible views when filming multiview content. View synthesis is performed using the available views and depth maps. There is a correlation between the quality of the synthesized views and the quality of depth maps. In this paper we study the effect of depth map quality on perceptual quality of synthesized view through subjective and objective analysis. Our evaluation results show that: 1) 3D video quality depends highly on the depth map quality and 2) the Visual Information Fidelity index computed between the reference and distorted depth maps has Pearson correlation ratio of 0.75 and Spearman rank order correlation coefficient of 0.67 with the subjective 3D video quality.

## 1. INTRODUCTION

The quality of multiview video systems has been improving in the recent years. Multiview systems provide viewers with realistic depth impression through 3D content, which includes two or more views of the same scene. Since it is not practical to capture and transmit many views of the same scene, only some of the views are captured using synchronized cameras, and the rest of required views are synthesized using the captured views. Figure 1 depicts the delivery pipeline of multiview content suggested by the Joint Collaborative Team on 3D Video Coding Extension Development (JCT-3V), which is a group of experts from ITU-T Study Group 16 (VCEG), and ISO/IEC JTC 1/SC 29/WG 11 (MPEG) [1]. As it can be seen, a limited number of captured views and their corresponding depth maps are encoded and transmitted as a bit stream. Then, at the receiver side these bit streams are decoded and several intermediate views are synthesized using the available views and their depth maps for multiview applications. Transmission of depth map information requires much less bandwidth compared to regular views, since depth map frames are gray scale images with minimal texture information. In the view synthesizing process, the quality of rendered views depends on the quality of available views as well as depth information.

It is well known that the quality of a depth map cannot be quantified by using perceptual-based distortion metrics, since depth maps are not natural images. Instead, one approach is to consider unique properties of the depth map sequence such as sparsity and study the impact of these properties on the 3D perceptual quality of the 3D views. Following this idea, the authors in [2] utilized the temporal and spatial variances of the disparity map to define a metric for measuring the depth map distortions. Following the same concept, another method uses Euclidean distance between the reference and distorted depth maps to measure the depth map impairment level [3].

Unlike the previous approaches, other studies treat the depth maps as gray scale images. In that respect, depth map quality is measured using the perceptual-based image quality metrics. Following this idea different works utilize SSIM (Structural Similarity) or VIF (Visual Information Fidelity) indexes to quantify the quality of the depth map and incorporate its effect in the design of their proposed 3D picture quality metrics [4 – 7]. In this paper we investigate the relationship between the overall 3D video quality of experience and the depth map quality measured by perceptual-based image quality metrics. To this end, several particularly probable depth map artifacts are simulated and applied to the depth map sequence, and the synthesized views are generated using the original views and the distorted depth map sequence. The quality of the depth map is measured using perpetual-based image quality metrics and its relationship with the subjective quality results of the synthesized stereo videos is studied.

The rest of this paper is organized as follows: Section 2 provides details on the setup of our experiment, experiment results and discussions are provided in Section 3, and Section 4 concludes the paper.

## 2. EXPERIMENT SETUP

To study the effect that depth map quality has on the perceived 3D video quality, first we apply common probable depth map artifacts to the depth map sequence corresponding to left view of a stereo video pair. Then the right view of the stereo video pair is synthesized using the left view and its corresponding modified depth map sequence. Note that the left view utilized to synthesize the right view has original quality (it has not been distorted by any kind of processing or distortions). Later we measure the correlation between the quality of depth map sequences and the overall subjective 3D Quality of Experience (QoE). The following subsections elaborate on our experiment setup and subjective test procedure.

### 2.1 Dataset

Five original stereo video sequences are selected from the test set recommended by the MPEG for Call for Proposals on 3D Video Coding Technology [8] for our experiment. The properties of these sequences are given in Table I and the snapshots of the left views of each stereo pair are demonstrated in Figure 2.

This work was partly supported by Natural Sciences and Engineering Research Council of Canada (NSERC) and the Institute for Computing Information and Cognitive Systems (ICICS) at UBC.

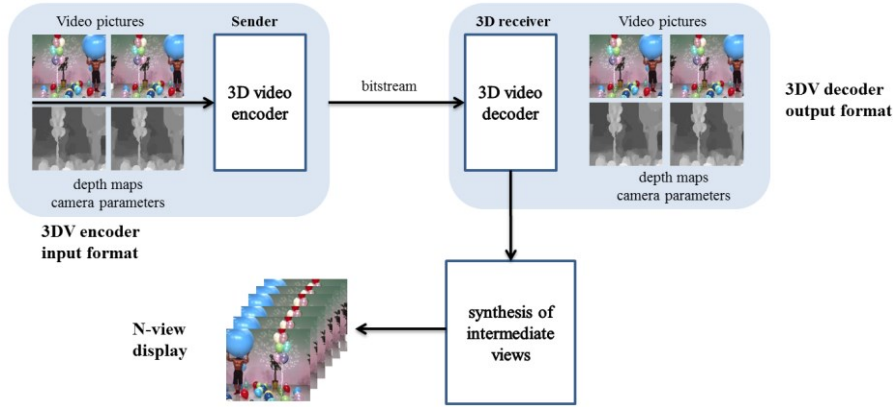

Figure 1. Multiview content delivery pipeline

Note that for each view of the stereo pairs, the corresponding depth map sequence is available.

The most possible artifacts that can degrade the depth map quality within the 3D video transmission pipeline (see Figure 1) are generated due to compression and packet loss. In our study, in order to simulate the compression artifacts we compress the depth map sequences corresponding to the left views in the dataset using the High Efficiency Video Coding (HEVC) standard (HM software version 9 [9]) with Quantization Parameters (QPs) of 25 and 45, and low delay profile with GOP size of 4. In order to simulate the packet loss effect, the H.264/AVC-based network transmission simulator function in JM software, version 16.2, is utilized. Packet loss density was set to 3% and 10%.

After applying the compression and packet loss artifacts to the depth map sequences, the right view is synthesized using the available left view and the distorted depth map sequence corresponding to the left view. The MPEG View Synthesis Reference Software (VSRS version 3.5 [10]) was used to generate the synthesized view.

### 2.2 Subjective tests procedure

Once the synthesized views are generated, for each video sequence, the right synthesized view is paired with the original left view to form the test stereo video. Considering that there are five original videos with four different distortions (two resulted from the 2 levels of compression and two from the 2 levels of packet loss), the entire test dataset included 20 video streams.

Fifteen male/female subjects with age between 25 and 32 years old participated in our tests. All of them were screened for color and visual acuity (using Ishihara and Snellen charts), and also for stereo vision (Randot test – graded circle test 100 seconds of arc). The evaluation was performed using a 46" Full HD Hyundai 3D TV (Model: S465D) with passive glasses. The TV settings were as follows: brightness: 80, contrast: 80, color: 50, R: 70, G: 45, B: 30. The 3D display and the settings were based on the MPEG recommendations for the subjective evaluation of the proposals submitted in response to the 3D Video Coding Call for Proposals [8].

The subjective tests were carried out according to the ITU BT.500-13 standard using a single stimulus method for the test sessions and an "adjectival categorical judgment method" for the ratings [11]. Each test session was designed according to the single stimulus method, such that the videos of the same scene with different impairments were shown to the observers in a random order, and the subjects were rating each and every video. There were five discrete quality levels (1-5: bad, poor, fair, good, excellent) for rating the videos, where score 5 indicated the highest quality and score 1 indicated the lowest quality. In particular, subjects were asked to rate a combination of "naturalness", "depth impression" and "comfort" as suggested by [12]. Note that during the test, the videos of the same scene with dif-ferent qualities were presented in a single random order to each subject (the order was not changed for different subjects).

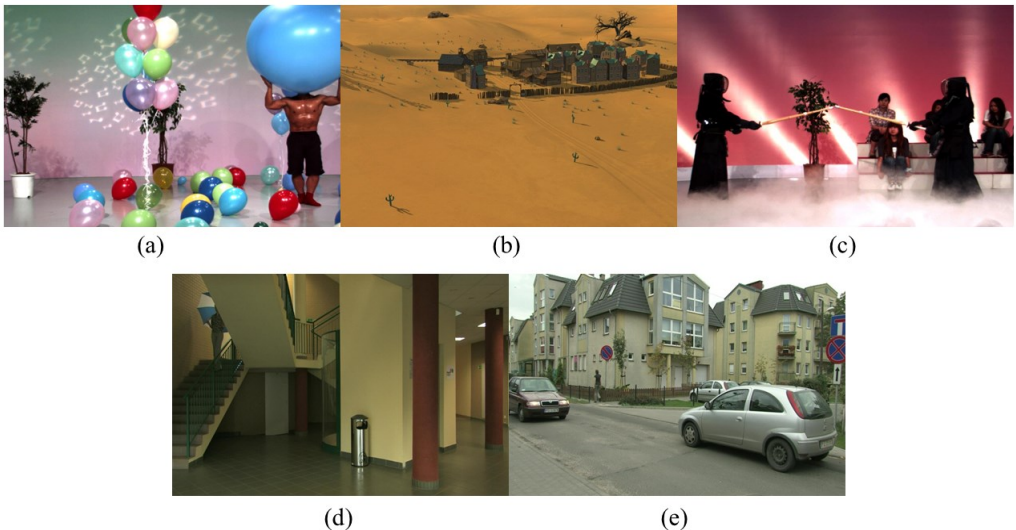

Figure 2. Snapshots of test videos

TABLE I: TEST VIDEOS

| Test Sequence | Resolution | Frame Rate (fps) | Number of Frames | Spatial Complexity | Temporal Complexity | Input Views | View to synthesize | Stereo pair |
|---|---|---|---|---|---|---|---|---|
| Poznan_Hall2 | 1920×1088 | 25 | 200 | Medium | Medium | 7-6 | 6.5 | 6.5-6 |
| Poznan_Street | 1920×1088 | 25 | 250 | High | High | 4-3 | 3.5 | 3.5-3 |
| GT_Fly | 1920×1088 | 25 | 250 | High | High | 5-2 | 4 | 4-2 |
| Kendo | 1024×768 | 30 | 400 | Medium | Medium | 3-5 | 4 | 4-5 |
| Balloons | 1024×768 | 30 | 500 | Medium | Medium | 3-5 | 4 | 4-5 |

After collecting the subjective test results, a rejection analysis was performed to detect the outliers. A subject is labeled as an outlier if the correlation between the Mean Opinion Score (MOS) and the subject's rating scores for all videos is less than 0.75. We found that there was no outlier among the subjects.

## 3. RESULTS AND DISCUSSIONS

In order to get the objective evaluation of the quality of the depth maps, we compared the reference depth map videos with the distorted depth map videos using various perceptual quality metrics. To evaluate the quality of distorted depth maps in our experiment, we used the following quality metrics: PSNR, SSIM [5], MS-SSIM (Multi Scale SSIM [13]), DCT-based video quality metric (VQM) [14], and VIF [7]. Figure 3 illustrates the relation between the perceptual quality of stereo views and the quality of the depth map using different quality metrics. A logistic curve fitting has been applied to each set of results as follows [15]:

$$y = \frac{a}{1 + e^{-b(x-c)}} \quad (1)$$

where $x$ denotes the horizontal axis and $y$ represents the vertical axis in each diagram. Note that the MOS values in Figure 3 are normalized to unity. As it can be observed, VIF shows better correlation with MOS than all the other metrics.

In order to statistically measure the performance of each mapping we calculated the Spearman rank order correlation coefficient (to measure the monotonicity/reliability), the Pearson correlation ratio and Root Mean Square Error (RMSE) (to measure the accuracy), and the outlier ratio (to measure the consistency). Table II, Table III, and Table IV illustrate the statistical dependency of the quality metrics used in our study with the subjective tests. We can observe that the overall perceived 3D video quality highly depends on the depth map quality. Also, as already mentioned above, by comparing the performance of the quality metrics we observe that the Visual Information Fidelity has the highest correlation with MOS, for the compression artifact, packet loss artifact, and also for both artifacts. VIF and also other metrics in Table II, III, and IV are originally proposed as quality metrics for natural images. Our study shows that they have also a fair performance for depth map quality evaluation. This might be because depth maps roughly include some portions of the scene structure. The reported correlation values in Table IV also confirm that depth map alone is not sufficient for predicting the overall 3D video quality. In addition to depth map quality, 3D video quality depends on other factors such as the quality of the individual views and quality of the cyclopean view [6].

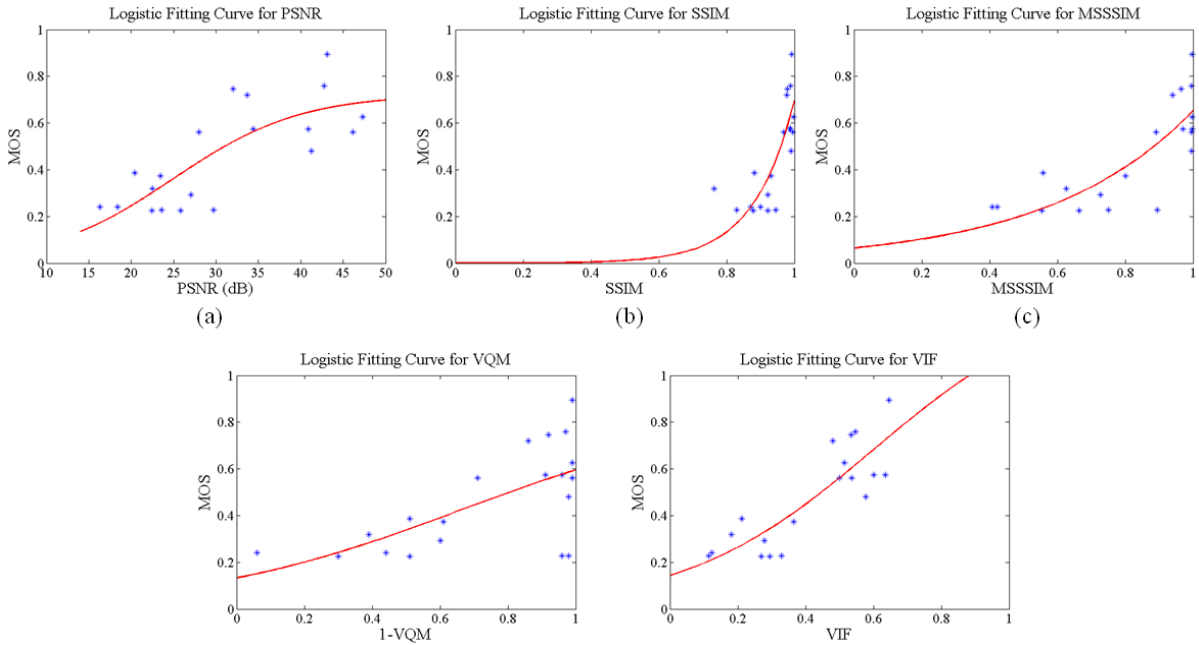

Figure 3. Comparing the subjective results with objective results using PSNR (a), SSIM (b), MS-SSIM (c), VQM (d), and VIF (e) objective quality metrics

TABLE II. CORRELATION BETWEEN THE SUBJECTIVE TESTS AND QUALITY METRICS, COMPRESSION ARTIFACT

| Quality Metric | Spearman Ratio | Pearson Ratio | RMSE | Outlier Ratio |
|---|---|---|---|---|
| PSNR | 0.7146 | 0.7651 | 0.1103 | 0 |
| SSIM | 0.6842 | 0.6668 | 0.1253 | 0 |
| MS-SSIM | 0.6660 | 0.7077 | 0.1229 | 0 |
| VQM | 0.6522 | 0.6623 | 0.1623 | 0 |
| **VIF** | **0.7997** | **0.7998** | **0.0971** | **0** |

TABLE III. CORRELATION BETWEEN THE SUBJECTIVE TESTS AND QUALITY METRICS, PACKET LOSS ARTIFACT

| Quality Metric | Spearman Ratio | Pearson Ratio | RMSE | Outlier Ratio |
|---|---|---|---|---|
| PSNR | 0.6195 | 0.4756 | 0.1384 | 0 |
| SSIM | 0.6988 | 0.7259 | 0.1033 | 0 |
| MS-SSIM | 0.6866 | 0.6795 | 0.1164 | 0 |
| VQM | 0.5915 | 0.6067 | 0.1478 | 0 |
| **VIF** | **0.7110** | **0.7585** | **0.0795** | **0** |

TABLE IV. CORRELATION BETWEEN THE SUBJECTIVE TESTS AND QUALITY METRICS, THE WHOLE VIDEO SET

| Quality Metric | Spearman Ratio | Pearson Ratio | RMSE | Outlier Ratio |
|---|---|---|---|---|
| PSNR | 0.6209 | 0.6535 | 0.1324 | 0 |
| SSIM | 0.6420 | 0.6126 | 0.1247 | 0 |
| MS-SSIM | 0.6541 | 0.6623 | 0.1242 | 0 |
| VQM | 0.5748 | 0.5361 | 0.1602 | 0 |
| **VIF** | **0.6706** | **0.7451** | **0.1086** | **0** |

In designing an efficient coding scheme for multiview and depth map streams, quantifying the quality of depth map is important due to its correlation with the quality of synthesized views. The reported results in Table II help with selecting a proper quality metric for depth maps and adjusting the encoding parameters accordingly.

## 4. CONCLUSION

In this paper we studied the performance of perceptual quality metrics in measuring the quality depth map videos in the context of 3D video transmission. Two possible depth map artifacts were simulated and applied to depth map sequences. The perceptual quality of the synthesized stereo videos generated based on the distorted depth maps was tested. Subjective evaluations showed that the depth map quality highly correlates with the overall 3D quality. Moreover, objective quality assessment results showed that the Visual Information Fidelity (VIF) index calculated between the reference and distorted depth maps demonstrates the highest correlation with the 3D video Quality of Experience (QoE) compared to other perceptual-based quality metrics.